\documentclass[twocolumn,preprintnumbers, amssymb,amsmath,aps,floatfix,prl,nofootinbib,superscriptaddress,showpacs]{revtex4-1}

\usepackage{graphicx}
\usepackage{amsmath,amsthm,amssymb}

\begin{document}

\preprint{YITP-14-93}
\title{Analytical and numerical Gubser solutions of the second-order hydrodynamics}

\author{Long-Gang Pang}
\affiliation{Key Laboratory of Quark and Lepton Physics (MOE) and Institute
of Particle Physics, Central China Normal University, Wuhan 430079, China}

\author{Yoshitaka Hatta}
\affiliation{Yukawa Institute for Theoretical Physics, Kyoto University, Kyoto 606-8502, Japan}

\author{Xin-Nian Wang}

\affiliation{Key Laboratory of Quark and Lepton Physics (MOE) and Institute
of Particle Physics, Central China Normal University, Wuhan 430079, China}
\affiliation{Nuclear Science Division Mailstop 70R0319, Lawrence Berkeley National
Laboratory, Berkeley, California 94720, USA}

\author{Bo-Wen Xiao}
\affiliation{Key Laboratory of Quark and Lepton Physics (MOE) and Institute
of Particle Physics, Central China Normal University, Wuhan 430079, China}

\begin{abstract}

Evolution of quark-gluon plasma near equilibrium can be described by the second-order relativistic viscous hydrodynamic equations. Consistent and analytically verifiable numerical solutions are critical for phenomenological studies of the collective behavior of quark-gluon plasma in high-energy heavy-ion collisions. A novel analytical solution based on the conformal Gubser flow that is a boost-invariant solution with transverse fluid velocity is presented. Because of the nonlinear nature of the equation, the analytical solution is nonperturbative and exhibits features that are rather distinct from solutions to usual linear hydrodynamic equations. It is used to verify with high precision the numerical solution with a newly developed state-of-the-art $(3+1)$-dimensional second-order viscous hydro code (CLVisc). The perfect agreement between the analytical and numerical solutions demonstrates the reliability of the numerical simulations with the second-order viscous corrections. This lays the foundation for future phenomenological studies that allow one to gain access to the second-order transport coefficients.

\end{abstract}
\maketitle

\section{Introduction}
Relativistic hydrodynamics has been one of the essential tools to study the properties of the quark-gluon plasma (QGP) created in ultrarelativistic heavy-ion collisions \cite{Heinz:2013th}. A picture of the QGP as a nearly perfect fluid emerged from comparisons between experimental data and viscous hydrodynamic simulations \cite{Luzum:2008cw} with a small specific shear viscosity (shear viscosity to entropy ratio $\eta_v/s$) that is very close to the lower bound  $1/4\pi$ \cite{Kovtun:2004de} computed for $\mathcal{N}=4$ super-Yang-Mills (SYM) theory in the AdS/CFT correspondence. The extraction of the specific shear viscosity relies on numerical solutions of the viscous hydrodynamics with realistic initial conditions.

There has been tremendous progress in solving relativistic ideal and viscous hydrodynamic equations numerically with realistic initial conditions to simulate the dynamical evolution of the dense matter in heavy-ion collisions \cite{Kolb:2000sd, Rischke:1995ir, Hirano:2001eu, Petersen:2008dd, Werner:2010aa, Romatschke:2007mq, Song:2007ux, Schenke:2010rr, Schenke:2010nt, Bozek:2011ua, Nonaka:2013uaa, Karpenko:2013wva, Shen:2014vra, Molnar:2014zha}.
Analytical solutions, even with simplified initial conditions,  can also play a very important role in understanding the evolution dynamics and testing the consistency of numerical solutions. Bjorken flow \cite{Bjorken:1982qr} is a well-known analytical solution to the ideal hydrodynamic equation for a transversely uniform and longitudinally boost-invariant system. It has been recently extended to the Gubser flow \cite{Gubser:2010ze, Gubser:2010ui} by including nontrivial transverse flow velocity with the help of conformal symmetry. Moreover, an exact solution to the first-order viscous hydrodynamic equation (the Navier-Stokes equation), which reduces to the Gubser flow in the ideal limit, was also found \cite{Gubser:2010ze, Gubser:2010ui}.
On general grounds, one expects that the relativistic Navier-Stokes equation is pathological, and indeed this solution \cite{Gubser:2010ze, Gubser:2010ui} shows unphysical behaviors such as a negative temperature at early time. Though attempts have been made to cure this problem by solving, semianalytically and numerically, the Israel-Stewart equation  \cite{Marrochio:2013wla} and the microscopic Boltzmann equation in the relaxation time approximation \cite{Denicol:2014xca, Denicol:2014tha, Nopoush:2014qba}, it is important to search for more complete and consistent second-order relativistic hydrodynamic equations \cite{DeGroot:1980dk,Koide:2006ef,Baier:2007ix,Bhattacharyya:2008jc, Natsuume:2007ty,PeraltaRamos:2009kg,Denicol:2012cn,Tsumura:2012kp} 
 and their solutions.

In this paper,  we will go beyond the Israel-Stewart equation and find an exact and well-behaved analytical solution of the conformal second-order hydrodynamic equation from Ref.~\cite{Baier:2007ix} that reduces to the Gubser flow in a certain limit.  We furthermore will use this analytical solution to check the accuracy of numerical solutions of the complete second-order viscous hydrodynamic equation based on CCNU-LBNL viscous hydrodynamic model (CLVisc) and in turn test numerically the stability of the analytic solution. Using CLVisc and proper initial conditions, we find almost perfect agreement with the analytical solution within the accuracy of the numerical simulations. We also find that small perturbations to initial conditions dissipate quickly after a few $\textrm{fm/c}$ of hydrodynamical evolution, indicating the stability of the solution. The study can help to build the state-of-art and analytically tested second-order viscous hydrodynamic models, and eventually lead us to a better understanding of second-order transport coefficients for the QGP in future phenomenological studies.

\section{The analytical solution to the conformal hydrodynamic equation}

We work in  the $(\tau,x,y,\eta)$ or $\tau-\eta$ coordinates for which $\tau$ is the proper time and $\eta$ is the spatial rapidity. The metric in this coordinate system is
\begin{equation}
ds^2=d\tau^2-dx_\perp^2-x_\perp^2d\phi^2-\tau^2d\eta^2, \label{met}
\end{equation}
where $x_\perp=\sqrt{x^2+y^2}$. The second-order hydrodynamic equation without external currents is simply
given by
\begin{equation}
\nabla_{\mu} T^{\mu\nu} =0 \label{t0},
\end{equation}
with the energy-momentum tensor $T^{\mu\nu}=\epsilon u^\mu u^\nu -p \Delta^{\mu\nu}+\pi^{\mu\nu} $, where $\epsilon$ is the energy density, $p$ the pressure, $u^\mu$ the flow 4-velocity normalized as $u^\mu u_\mu=1$, and $\Delta^{\mu\nu} =g^{\mu\nu}-u^\mu u^\nu$ the projection operator orthogonal to the flow velocity. The shear pressure tensor $\pi^{\mu\nu}$ represents the deviation from ideal hydrodynamics and local equilibrium. We choose to work in the Landau frame, which yields ta ransverse ($u_\mu \pi^{\mu\nu}=0$) and traceless ($\pi^\mu_{\ \,\mu} =0$) shear stress tensor. Our assumed conformal symmetry implies $T^\mu_{\ \,\mu} =0$ and the equation of state $\epsilon=3p$. By projecting along the flow velocity $u^\mu$ and the direction orthogonal to $u^\mu$, we can rewrite the hydrodynamic equation as,
\begin{eqnarray}
&& D \epsilon + (\epsilon +p )\vartheta-\frac{1}{2}\pi^{\mu\nu}\sigma_{\mu\nu}=0\,, \label{a} \\
&& (\epsilon +p ) D u^\mu - \Delta^{\mu\alpha}\nabla_\alpha p  + \Delta^\mu_{\ \nu}\nabla_\alpha \pi^{\alpha\nu}=0\,, \label{b}
\end{eqnarray}
respectively, where $D=u^\mu \nabla_\mu$ is the comoving derivative and $\vartheta=\nabla_\mu u^\mu$ the expansion rate. The traceless shear viscous pressure tensor $\pi^{\mu\nu}$ satisfies the  equation \cite{Baier:2007ix},
\begin{eqnarray}
\pi^{\mu\nu} &=&\eta_v \sigma^{\mu\nu}-\tau_\pi \left[ \Delta^\mu_\alpha\Delta^\nu_\beta u^\lambda \nabla_\lambda \pi^{\alpha\beta} +\frac{4}{3}\pi^{\mu\nu} \vartheta\right] \notag \\
&&- \lambda_1 \pi^{\langle\mu}_{\ \ \lambda} \pi^{\nu\rangle\lambda}-\lambda_2 \pi^{\langle \mu}_{\ \ \lambda} \Omega^{\nu\rangle\lambda}- \lambda_3 \Omega^{\langle \mu}_{\ \  \lambda}\Omega^{\nu\rangle \lambda},\label{pi}
\end{eqnarray}
with the symmetric shear tensor $\sigma^{\mu\nu}$ and the antisymmetric vorticity tensor $\Omega^{\mu\nu}$ defined as
\begin{eqnarray}
\sigma^{\mu\nu}&\equiv& 2\nabla^{\langle \mu} u^{\nu\rangle} \equiv 2 \Delta^{\mu\nu\alpha \beta} \nabla_\alpha u_\beta\, , \nonumber\\
\Omega^{\mu\nu}&\equiv& \frac{1}{2}\Delta^{\mu\alpha}\Delta^{\nu\beta}(\nabla_\alpha u_\beta -\nabla_\beta u_\alpha) ,\nonumber\\
\Delta^{\mu\nu\alpha \beta}&\equiv&\frac{1}{2}(\Delta^{\mu\alpha}\Delta^{\nu\beta}
+\Delta^{\mu\beta}\Delta^{\nu\alpha})-\frac{1}{3}\Delta^{\mu\nu}\Delta^{\alpha\beta} ,
\end{eqnarray}
where $\Delta^{\mu\nu\alpha \beta}$ is the double projection operator that renders the resulting contracted tensors symmetric, traceless and orthogonal to the flow velocity. In Eq.~(\ref{pi}), $\tau_\pi$, $\lambda_1$, $\lambda_2$, and $\lambda_3$ are four independent second-order transport coefficients in flat space-time.

Following Gubser \cite{Gubser:2010ze}, we perform a conformal/Weyl transformation to the coordinate system
\begin{equation}
d\hat{s}^2\equiv \frac{ds^2}{\tau^2}
= d\rho^2- \cosh^2\rho (d\theta^2+\sin^2\theta d\phi^2)- d\eta^2\,, \label{dsr}
\end{equation}
which indicates that the Minkowski space is conformal to $dS_3\times \mathcal{R}$  with,
\begin{equation}
\sinh \rho=-\frac{L^2-\tau^2+x_\perp^2}{2L\tau}, \quad \tan \theta=\frac{2L x_\perp}{L^2+\tau^2-x_\perp^2},
\end{equation}
where $L$ can be interpreted as the radius of the $dS_3$ space and may be understood as the typical size of the relativistic fluid in phenomenology. Hereafter, dynamical variables in the new coordinates $\hat{x}^\mu=(\rho, \theta, \phi, \eta)$ will carry a hat to avoid confusion. The Gubser flow and our new solution are both characterized by the comoving flow velocity $ \hat u^\mu\equiv(1,0,0,0)$ in the $\hat{x}^\mu$ coordinates. It is straightforward to find that,
\begin{eqnarray}
&&\hat\vartheta=2\tanh \rho, \quad \,
\hat\Omega^{\mu\nu}=0, \nonumber \\
&&\hat\sigma^\theta_{\,\theta}=\hat\sigma^\phi_{\,\phi}=-\frac{1}{2}\hat \sigma^\eta_{\,\eta}=\frac{2}{3}\tanh \rho  \,.
\end{eqnarray}
We factor out various powers of $\epsilon$ from all the transport coefficients so that
$\hat\eta_v=\eta/\epsilon^{3/4}$, $\hat\tau_{\pi}=\tau_{\pi}\epsilon^{1/4}$, $\hat\lambda_1=\lambda_1\epsilon$, $\hat\lambda_2=\lambda_2 \epsilon^{1/4}$ and $\hat\lambda_3=\lambda_3/\epsilon^{1/2}$ are dimensionless.  After the conformal transform, Eq~(\ref{pi}) then becomes,\begin{eqnarray}
&&\hat\pi^{\mu\nu} =\hat\eta_v \hat\epsilon^{3/4}\hat\sigma^{\mu\nu}-\frac{\hat\tau_\pi}{\hat\epsilon^{1/4}}\left[ \hat\Delta^\mu_\alpha \hat \Delta^\nu_\beta \hat u^\lambda \hat\nabla_\lambda \hat\pi^{\alpha\beta} +\frac{4}{3}\hat\pi^{\mu\nu} \hat\vartheta\right] \notag \\
&& -\frac{\hat\lambda_1}{\hat\epsilon} \hat\pi^{\langle\mu}_{\ \ \lambda} \hat\pi^{\nu\rangle\lambda}
-\frac{\hat\lambda_2}{\hat\epsilon^{1/4}} \hat\pi^{\langle \mu}_{\ \ \lambda} \hat\Omega^{\nu\rangle\lambda}- \hat\lambda_3\hat\epsilon^{1/2} \hat\Omega^{\langle \mu}_{\ \  \lambda}\hat\Omega^{\nu\rangle \lambda}, \label{pi2}
\end{eqnarray}
where $\hat\epsilon=\epsilon\tau^4$ is also dimensionless. Assuming $\hat\pi^{\mu\nu}$ is diagonal and has the form $\hat\pi^{\mu\nu} =(0, \hat\pi^{\theta\theta}, \hat\pi^{\phi\phi}, \hat\pi^{\eta\eta})$, one can show that Eqs.~(\ref{a}) and (\ref{b}) can be cast into,
\begin{eqnarray}
&&\partial_\rho \hat\epsilon+\frac{8}{3}\hat\epsilon \tanh \rho -C\tanh\rho=0 \, , \\
&& \hat\pi^{\theta\theta}=\hat\pi^{\phi\phi}\sin^2\theta \, ,
\end{eqnarray}
respectively. In addition, Eq.~(\ref{pi2}) can be written as,
\begin{widetext}
\begin{eqnarray}
&&\left[\partial_\rho A+\frac{8}{3}A\tanh\rho+\frac{2}{3}\frac{\hat\eta_v \hat\epsilon}{\hat\tau_\pi}\tanh\rho\right]+\frac{\hat\epsilon^{1/4}}{\hat\tau_\pi}\left[A-\frac{\hat\lambda_1}{3\hat\epsilon}\left(2A^2-B^2-C^2\right)\right]=0,\\
&&\left[\partial_\rho B+\frac{8}{3}B\tanh\rho+\frac{2}{3}\frac{\hat\eta_v\hat \epsilon}{\hat\tau_\pi}\tanh\rho\right]+\frac{\hat\epsilon^{1/4}}{\hat\tau_\pi}\left[B-\frac{\hat\lambda_1}{3\hat\epsilon}\left(2B^2-C^2-A^2\right)\right]=0,\\
&&\left[\partial_\rho C+\frac{8}{3}C\tanh\rho-\frac{4}{3}\frac{\hat\eta_v\hat \epsilon}{\hat\tau_\pi}\tanh\rho\right]+\frac{\hat\epsilon^{1/4}}{\hat\tau_\pi}\left[C-\frac{\hat\lambda_1}{3\hat\epsilon}\left(2C^2-A^2-B^2\right)\right]=0, \label{pi4}
\end{eqnarray}
where $A\equiv\hat\pi^{\theta\theta}\cosh^2\rho$,  $B\equiv\hat\pi^{\phi\phi}\cosh^2\rho\sin^2\theta$ and $C\equiv\hat\pi^{\eta\eta}$. The above equations are a set of nonlinear differential equations, which are notoriously hard to solve analytically. Fortunately, when $\hat\eta_v\hat\lambda_1^2=3\hat\tau_\pi$, we manage to find a very simple analytical solution,
\begin{equation}
C=-2A=-2B=\frac{2}{\hat\lambda_1}\hat\epsilon \,  , \quad \textrm{and} \quad \hat\epsilon\propto \left(\frac{1}{\cosh\rho} \right)^{\frac{8}{3}-\frac{2}{\hat\lambda_1}}.
\end{equation}
After the Weyl rescaling, we can get back to the Minkowski $(\tau, x, y, \eta)$ space and obtain
\begin{eqnarray}
&&u_\mu = \tau \frac{\partial \hat{x}^\nu}{\partial x^\mu} \hat{u}_\nu =  \left[\frac{L^2+\tau^2+x_\perp^2}{\sqrt{(L^2+\tau^2+x_\perp^2)^2-4\tau^2x_\perp^2}}, \frac{-2\tau \vec{x}_\perp}{\sqrt{(L^2+\tau^2+x_\perp^2)^2-4\tau^2x_\perp^2}},0\right]\, ,\\
&&\epsilon=\frac{1}{\tau^4} \hat\epsilon  \quad \textrm{and} \quad \pi_{\mu\nu} = \frac{1}{\tau^2} \frac{\partial \hat{x}^\alpha}{\partial x^\mu} \frac{\partial \hat{x}^\beta}{\partial x^\nu}\hat{\pi}_{\alpha\beta}.
\end{eqnarray}
This conditional solution is very nontrivial since it involves three different transport coefficients and many nonvanishing components of the shear stress tensor $\pi_{\mu\nu}$. It is also useful for verifying numerical solutions of the second-order viscous hydrodynamic equations.
Our solution has the same transverse flow velocity  $v_\perp\equiv- u_\perp/u_\tau$ as the Gubser flow due to conformal symmetry.  In contrast to the pathological solution of the Navier-Stokes equation, which has negative temperature at early time, our second-order viscous solution is always well defined in the whole space-time.

The most interesting feature of the conformal second-order hydrodynamics studied in this paper is the nonlinearity of the equation. Since $\hat\lambda_1$ is found to be nonzero in both strongly and weakly coupled systems, it is important to consider the possibility that the conventional idea of perturbative solutions to linear hydro evolution equations could break down. Then, one has to include the nonlinear term $\lambda_1 \pi^{\langle\mu}_{\ \ \lambda} \pi^{\nu\rangle\lambda}$ and explore the unique structure of fixed-point solutions in nonlinear hydrodynamic equations.  Without  the nonlinear term $\lambda_1 \pi^{\langle\mu}_{\ \ \lambda} \pi^{\nu\rangle\lambda}$, the hydrodynamic equation is a set of linear differential equations in terms of $\pi^{\mu\nu}$. However, the nonlinear term can completely change the nature of the evolution and admit nonperturbative solutions such as the one found above. This is clear from the fact $\hat\pi^{\mu\nu} \sim \frac{1}{\lambda_1} \hat{\epsilon}$. Qualitatively speaking, this nonperturbative solution comes from the stable fixed point of nonlinear differential equations, which indicates that one cannot recover the solution to the Israel-Stewart equation simply taking $\hat\lambda_1\to 0$ limit for this solution. Furthermore, our numerical study presented below suggests that this fixed-point solution is stable. One more interesting observation is that the location of the fixed point approaches zero ($\pi^{\mu\nu}\to 0$) when we take $\hat\lambda_1\to \infty$, which allows us to recover the ideal solution. These are the truly distinct features of the nonlinear equation as compared to the linear hydrodynamic equation.

We note that the relation between $\hat\eta_v$, $\hat\lambda_1$ and $\hat\tau_\pi$ is designed to reduce the coupled differential equation with our analytic solution. It therefore does not provide any additional insights into the physical value of $\hat\lambda_1$. Such a relation only serves to find an analytic nonperturbative solution at a fixed point to a nonlinear hydro equation. Nevertheless, we conjecture that there could be a similar fixed point solution even when $\hat\lambda_1^2= 3 \hat\tau_\pi/\hat\eta_v$ is not satisfied. Therefore, the implication of this solution could be more general. Our study can provide some insight for understanding the nonlinearity of the second-order hydrodynamics and future phenomenological studies of the second-order transport coefficients.

\end{widetext}

\begin{figure}[tbp]
\begin{center}
\includegraphics[width=8cm]{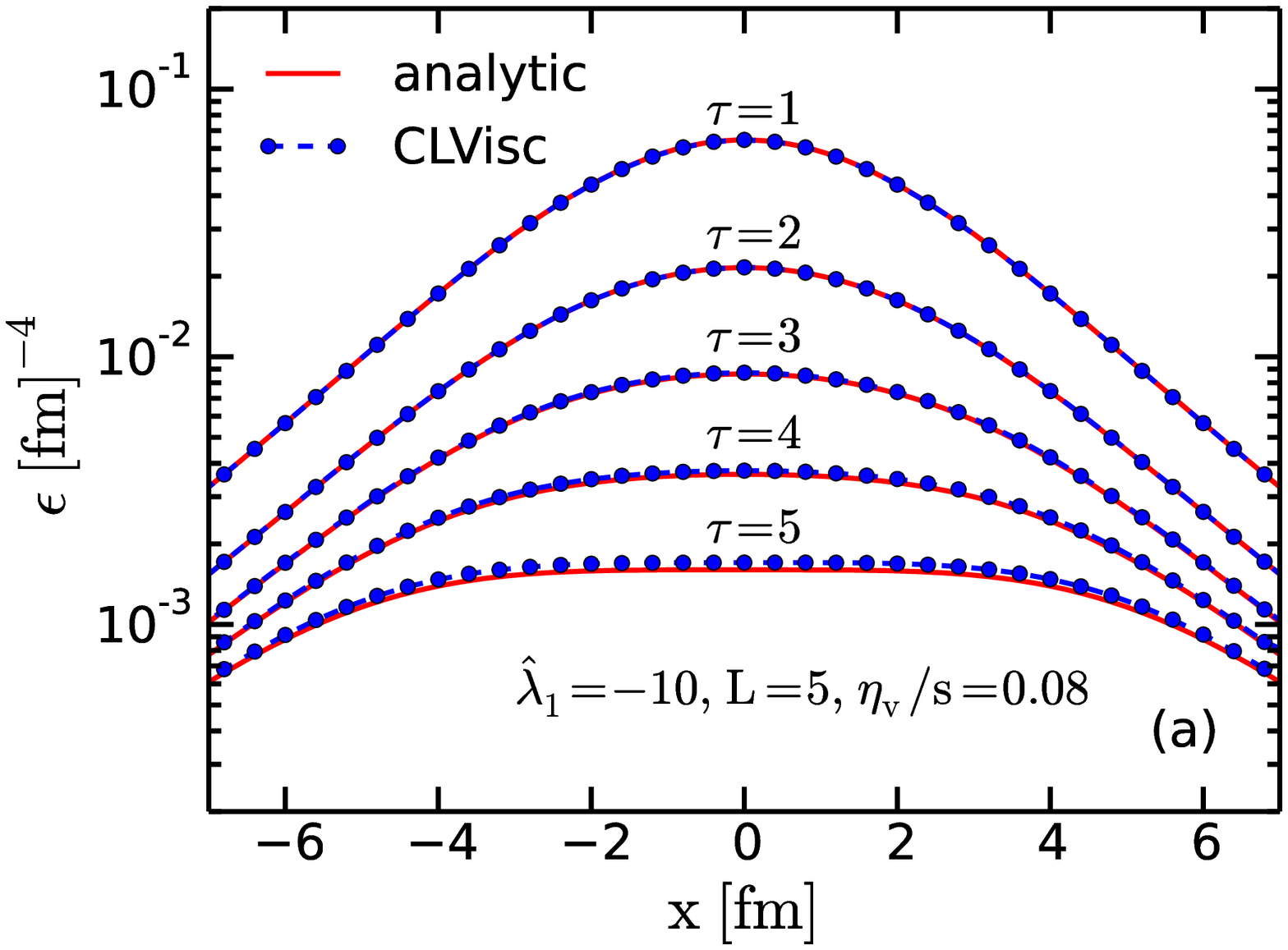}\\
\vspace{-5pt}
 \includegraphics[width=8cm]{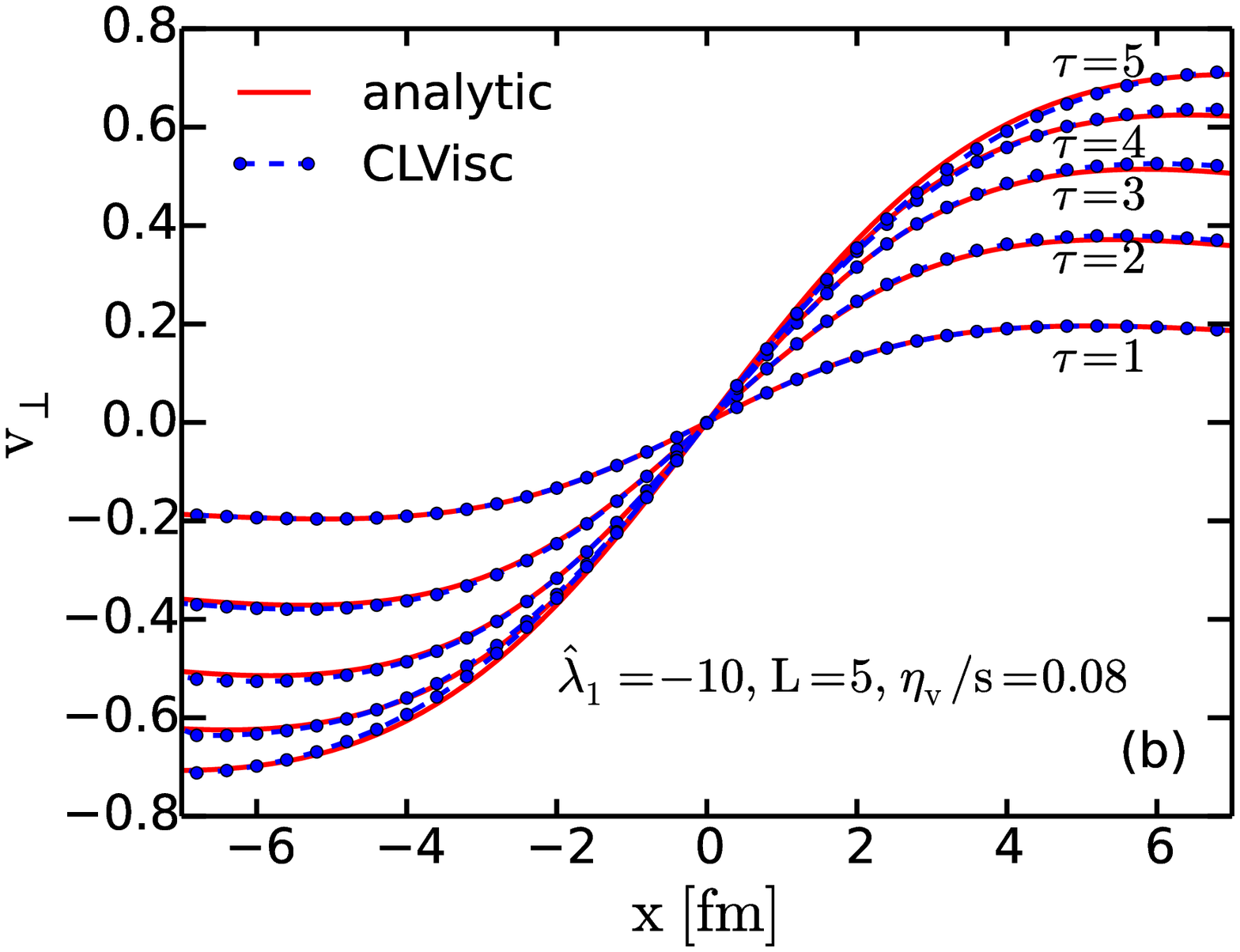}
\end{center}
\vspace{-16pt}
\caption[*]{Analytical and numerical results for (a) the energy density and (b) transverse flow velocity $v_\perp$ at $y=0$.}
\label{ev}
\end{figure}

For consistency and stability, the above solution is meaningful when $|\hat{\lambda_1}|\sim \hat{\epsilon}/|\hat{\pi}^\mu_\nu|\gg 1$.
For positive $\lambda_1$, we always get positive $\pi^{\eta\eta}$ and negative $\pi^{xx}\, , \pi^{yy}$. For negative $\lambda_1$, $\pi^{\eta\eta}$ becomes negative, while $\pi^{xx}\, , \pi^{yy}$ turn positive. In principle, $\lambda_1$ can be either positive or negative. A positive value ($\hat{\lambda}_1=3/4$) was reported for $\mathcal{N}=4$ super-Yang-Mills theory  in Ref.~\cite{Baier:2007ix}, whereas $\lambda_1$ is negative in a particular model considered in Ref.~\cite{Molnar:2013lta}. Since physical initial conditions for QGP in heavy-ion collisions lean toward positive $\pi^{xx}$ and $\pi^{yy}$, we shall employ a negative value for  $\hat{\lambda}_1$.

\begin{figure}[tbp]
\begin{center}
\includegraphics[width=8cm]{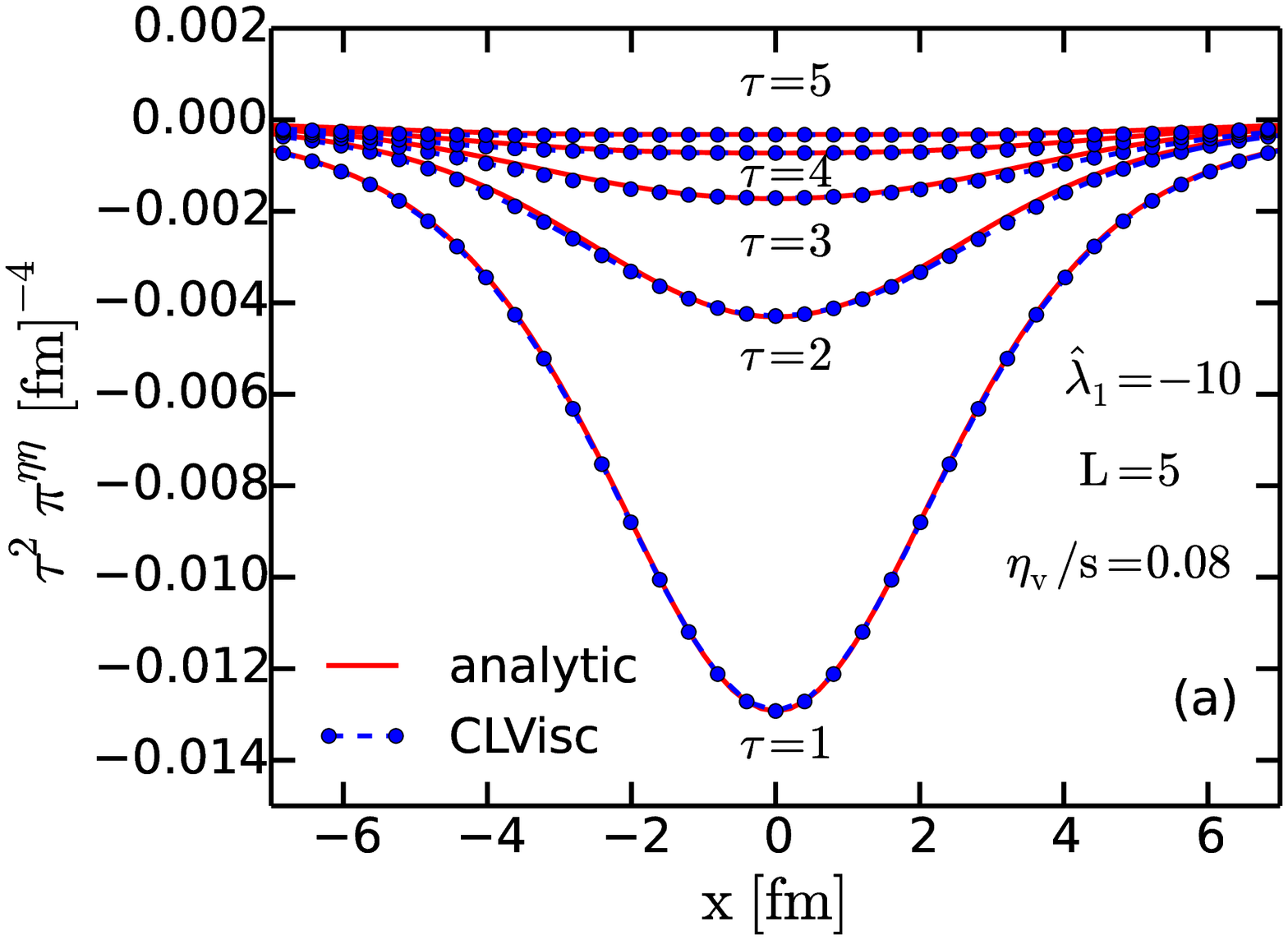}\\
\vspace{-5pt}
 \includegraphics[width=8cm]{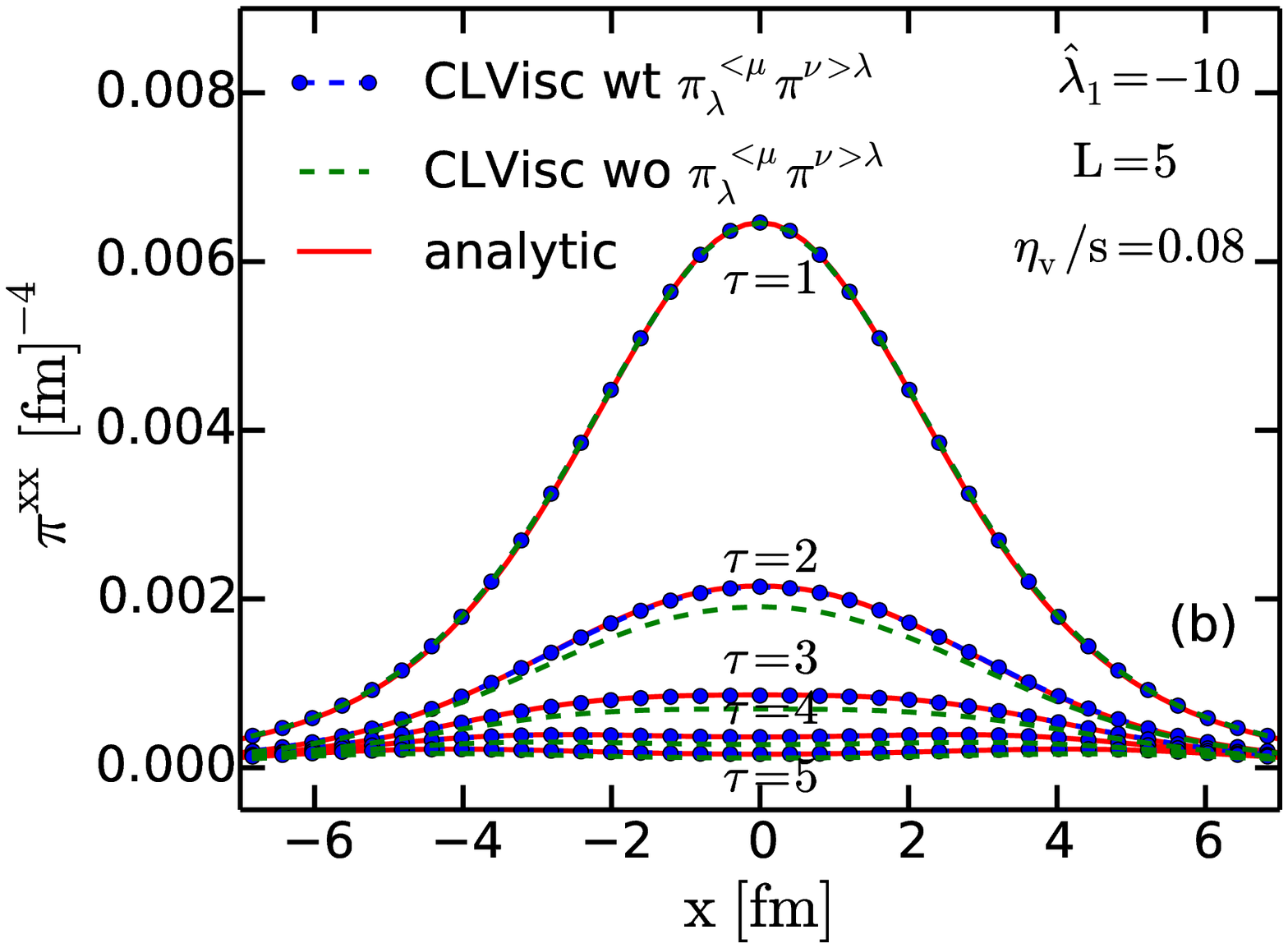}\\
 \vspace{-5pt}
\includegraphics[width=8cm]{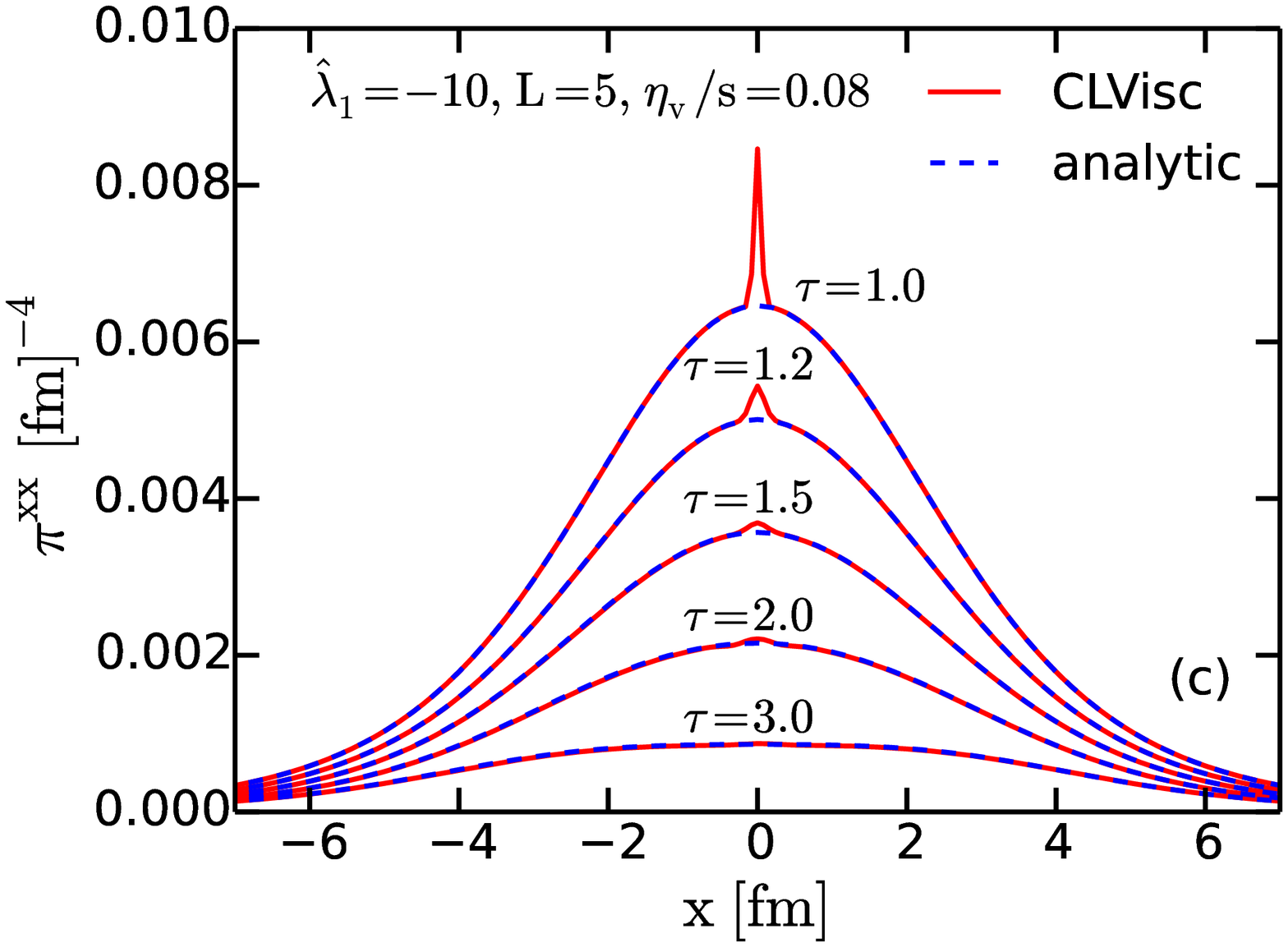}
\end{center}
\vspace{-16pt}
\caption[*]{Analytical and numerical results for (a) $\tau^2\pi^{\eta\eta}$,  (b) $\pi^{xx}$ at $y=0$ and (c) comparison to numerical results for $\pi^{xx}$ with an initial Gaussian perturbation.}
\label{shear}
\end{figure}

\section{Numerical results}

CLVisc is the extension of the ideal 3+1-dimensional hydrodynamic model  \cite{Pang:2012he} that includes the second-order viscous terms.  To solve Eqs.~(\ref{t0}) and (\ref{pi}), the original module for SHASTA algorithm \cite{Shasta} used in the ideal hydro code is replaced by the second-order central scheme Kurganov-Tadmor (KT) algorithm \cite{Kurganov:2002}, which is shown to be more stable in the low-energy density and high fluid velocity region. We employ the OpenCL GPU parallel language together with the KT algorithm implemented on graphic cards and the new code can reduce the simulation time by a factor of $10$ on a single GPU. 
CLVisc treats the shear stress tensor $\pi^{\mu\nu}$ as the source term of the ideal hydrodynamic energy-momentum tensor $T_{0}^{\mu\nu}=\epsilon u^\mu u^\nu -p \Delta^{\mu\nu}$ and implements them into two different evolution kernels, which makes the switch to ideal hydrodynamics easy. These two kernels are compiled on GPUs that process more efficiently most of the heavy computations such as KT evolutions and gradient calculation. At any given time, we can extract the energy density $\epsilon$ and flow velocity $u^\mu$ from $T_{0}^{\mu\nu}$.

Using CLVisc, we numerically solve the second-order viscous hydrodynamic equations for  $T^{\mu\nu}$ and $\pi^{\mu\nu}$ with the conformal equation of state $\epsilon=3p$ directly in Minkowski space-time in the $\tau$-$\eta$ coordinates. For later comparison, we set the initial condition at $\tau_0=1\, \textrm{fm/c}$ to match the analytical solution. We can then obtain $T^{\mu\nu}$ and $\pi^{\mu\nu}$ numerically according to the hydrodynamic evolution at $\tau > \tau_0$. The initial energy density $\epsilon$, fluid velocity $u^{\mu}$ and shear viscous tensor $\pi^{\mu\nu}$ are discretized on a lattice with the number of grids $N_x\times N_y\times N_\eta = 303\times303\times6$ and the grid size $\Delta x = \Delta y = 0.08\, \textrm{fm}$ and $\Delta\eta=0.3$.
To generate the correct time derivatives for initial fluid velocities $\partial_{\tau} u^{\mu}$, we set initial conditions for two time steps $\tau_{-1}=0.99$ fm/c and $\tau_0=1.0$ fm/c with the time evolution step $\Delta \tau=0.01$ fm/c. Given the above numerical setup in the KT algorithm, the numerical error can be estimated to be about a few percent at $\tau=5\, \textrm{fm/c}$. We find that it helps to reduce numerical errors by using $\tilde{\pi}^{\eta\eta}=\tau^2 \pi^{\eta\eta}$ instead of $\pi^{\eta\eta}$ directly in the numerical simulations, since the numerical derivatives become tricky due to nonvanishing Christoffel symbols in the $\tau$-$\eta$ coordinates. It was found in Ref.~\cite{Marrochio:2013wla} that some adjustment to the flux limiter is necessary in order to describe the shear stress tensor $\pi^{\mu\nu}$ in the semianalytic Gubser flow solution of the Israel-Stewart theory. For the smooth initial condition in our solution, we find that there is no need to make any adjustment to the flux limiter.

To compare with the analytical solution, we also set  $\hat\eta_v\hat\lambda_1^2=3\hat\tau_\pi$ and choose the parameters $\hat\lambda_1=-10$, $L=5 \textrm{ fm}$, and $\eta_v/s=3\hat\eta_v/4=0.08$, where we used the temperature $T\equiv \epsilon^{1/4}$. Shown in Fig.~\ref{ev} are the energy density (a) and the transverse flow velocity (b) from the analytical (solid lines) and CLVisc numerical solutions (dotted lines) for different values of time up to $\tau=5 \,\textrm{fm/c}$. The accumulated numerical relative error is roughly $5$ \% at $\tau=5 \,\textrm{fm/c}$ for the chosen spatial grid and time-step size. One could reduce the numerical error by decreasing the spatial grid and time-step size, however, with increased computing time. In Figs.~\ref{shear}(a) and \ref{shear}(b), we plot two different components of the shear stress tensor. We observe a perfect agreement again between the numerical and analytical results for these quantities. Furthermore, we find that the numerical results without the $\lambda_1$ term,
 shown as the (green) dashed curves in Fig.~\ref{shear}(b), have sizable differences from the results with the $\lambda_1$ term in the second-order viscous corrections.
  In addition, we show in Fig.~\ref{shear}(c)  that small two-dimensional Gaussian perturbations initially added to $\pi^{\mu\nu}$ at the central point dissipate quickly and have no effect on the hydrodynamic evolution for the rest of the system. It implies that both our numerical and analytical solutions are stable with respect to small perturbations.
 Incidentally, we have checked that an equally good agreement is obtained when $\lambda_1$ is positive and large. In the limit $|\hat\lambda_1|\to \infty$, both analytical and numerical solutions approach the ideal Gubser flow for both flow velocity and energy density.

\section{Conclusion}
We have found an analytical Gubser flow solution to the second-order conformal hydrodynamic equation beyond the Israel-Stewart theory. We have also solved the same second-order hydrodynamic equation numerically using the newly developed CLVisc hydro code. The numerical solution agrees perfectly with the analytical one with the same condition and parameters. This gives us confidence in the numerical solutions beyond the Guber flow solution, at least for flows with vanishing vorticity. In the case with nonvanishing vorticity, it should be straightforward to follow the same procedure and compare with the analytical solution found earlier in Ref.~\cite{Hatta:2014gqa}. This paves the way for future phenomenological studies of QGP in heavy-ion collisions that can give us access to second-order transport coefficients.

\section*{Acknowledgments}
This work is supported by the NSFC under Grant No.~11221504, China MOST under Grant No. 2014DFG02050,  U.S. DOE under Contract No.~DE-AC02-05CH11231 and within the framework of the JET Collaboration. We thank J. Noronha for comments and discussions.

\end{document}